
\magnification=1200
\parindent 0pt
\parskip 5pt

\input epsf

\font\titlef=cmbx12 scaled\magstephalf
\font\sectf=cmbx10
\font\subsectf=cmsl10
\font\authorf=cmbx10
\font\affont=cmsl10

\font\figfont=cmbx6


\def\title#1#2{{\centerline{\titlef{#1}}}{\centerline{\titlef{#2}}}
\vskip .5cm}
\def\author#1{{\centerline{\authorf{#1}}}\medskip}
\def\affiliation#1#2#3{{\centerline{\affont{#1}}}{\centerline{\affont{#2}}}
{\centerline{\affont{#3}}}\vskip .5cm}
\def\abstract#1{{\bigskip\centerline{\bf ABSTRACT}}\medskip{#1}
}

\newcount\sectionnumber
\newcount\eqnum 

\newcount\subsectionnumber
\newcount\fignumber

\def\section#1{\global\subsectionnumber=0\global\eqnum=0
\medskip\global\advance\sectionnumber by 1 
${\hbox{\bf{\the\sectionnumber.}}}\,\,{{\hbox{\sectf{#1}}}}$\medskip}   
\def\subsection#1{\medskip \global\advance\subsectionnumber by 1
${\hbox{\sl{\the\sectionnumber-\the\subsectionnumber}}}\,\,
{\underline{\hbox{\subsectf{#1}}}}$\medskip}

\def\appendix#1{\global\aeqnum=0
\bigskip\global\advance\appnumber by 1
\centerline{${\hbox{{\bf Appendix}}\,\,{\hbox{\bf{\the\appnumber}}}}$}
\centerline{#1}
\bigskip}


\def\num{(\the\sectionnumber.\the\eqnum)}
\def\appnum{({\sl{A}}\the\appnumber.\the\aeqnum)}


\def\eqnu#1{\global\advance\eqnum by 1\eqno
{\scriptstyle\mit(\the\sectionnumber.\the\eqnum)}}

\def\aeqnu#1{\global\advance\aeqnum by 1
\eqno{\scriptstyle{\mit(A\the\appnumber.\the\aeqnum)}}}




\def\label{\edef}


\def\ie{{\it{i.e.\/}}}
\def\eg{{\it{e.g.\/}}}

\def\M{{\cal{M}}}
\def\N{{\cal{N}}}
\def\B{{\cal{B}}}
\def\K{{\cal{K}}}
\def\Z{{\bf{Z}}}
\def\R{{\bf{R}}}
\def\S{{\cal{S}}}

\def\diver{{\rm{div}}\,}


\def\fig#1{\global\advance\fignumber by
1${\hbox{\figfont{Fig.\the\fignumber:}}}\,\,{{\hbox{\figfont{#1}}}}$}



\medskip
\hfill hep-th/0405087
\bigskip
\title{Torsion cycles as non-local magnetic sources}
{in non-orientable spaces} 
\author{Marcos Alvarez}
\affiliation{Centre for Mathematical Science, City University}
{Northampton Square, London EC1V 0HB, UK}{\tt m.alvarez@city.ac.uk}

\abstract{
Non-orientable spaces can appear to carry net magnetic charge, even in the
absence of magnetic sources. It is shown that this effect can be understood
as a physical manifestation of the existence of torsion cycles of codimension
one in the homology of space. 
}

\section{Introduction}

The idea that electromagnetic charge may be a manifestation of the topology
of space has a very long history [Whe, WM, Sor]. The basic configuration 
considered by Wheeler and Misner was a space that contained a handle-like 
region, often called a wormhole. This space is traversed by sourceless 
electric lines of force that enter the wormhole radially through one mouth 
and exit through the other one some distance away. For an observer situated 
sufficiently far from the wormhole, space appears approximately flat, but 
not sourceless: the two mouths look like two pointlike electric charges.

\bigskip
{\centerline{\epsfbox{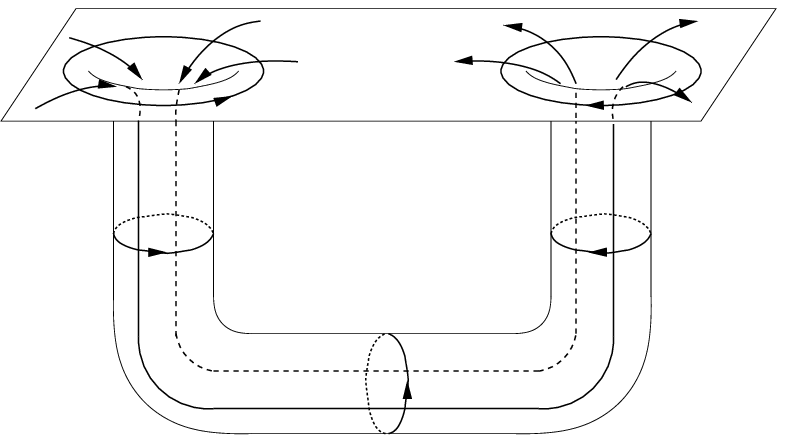}}\nobreak
{\centerline{\fig{An orientable handle}}}}
\bigskip

A ``dual'' construction can be imagined in which the lines of force are 
oriented circles that surround the wormhole mouths and extend to its interior
in a continuous way, as sketched in Fig.~1. From a distance, the two mouths
appear to carry dual charges of opposite signs. As a matter of terminology, we
will refer to the dual lines of force, and the charges they correspond to, as 
magnetic, although the motivation for doing that will only become clear later 
on. 

Serious objections can be raised against Misner and Wheeler's wormhole as a
realistic model of electromagnetic charge, on the grounds that wormholes are 
not satisfactory solutions in General Relativity. Under fairly general 
assumptions, the null energy condition has to be violated at some points near 
the wormhole throat [HV]. This indicates that wormhole solutions require the 
coupling of gravity to some type of exotic matter. Even if the right type of 
matter and coupling is found, known wormhole solutions are very unstable. 
Charged wormholes also exist in General Relativity in the presence of scalar 
fields non-minimally coupled to gravity. The scalar field provides a good 
example of the type of exotic matter required, but these charged wormholes are 
still unstable under spherically symmetric perturbations [BG]. 

Leaving aside these objections to the Misner-Wheeler idea of charge, the
strength of either of the two apparent charges would be defined by the flux 
through a sphere $\S$ that surrounds one of the mouths of the wormhole. The 
size of the sphere is of course immaterial, because the underlying field is 
divergenceless, and all such spheres are crossed by the same amount of flux 
and hence define the same charge strength. On the other hand, $\S$ has to be
large enough to be recognisably outside the wormhole throat. Unfortunately, the
construction given in [Whe, MW] lacks an intrinsic distinction between the part 
of space that belongs to the wormhole, and that part that remains ``outside'' 
of it. There is no clear way to tell when the sphere $\S$ is far away enough 
for its flux to be unambiguously attributable to one of the mouths of the 
wormhole. These ambiguities are due to the wormhole in [Whe, MW] being regarded 
as a purely topological feature of space, lacking any geometric structure. 

Another insatisfactory feature of the Misner-Wheeler construction is that it
is unable to accommodate the idea of an individual isolated charge. Any such
charge has to be understood as half of a dipole, the other half of which is
removed far away from it. Besides, the throat that joins the two apparent 
charges establishes a correlation between them that seems difficult to 
reconcile with quantum mechanics [Sor]. For these reasons, the 
Misner-Wheeler wormhole can never be taken seriously as a realistic model of 
charged particles.

Some time later, Sorkin constructed a non-orientable version of the Wheeler and 
Misner wormhole, and showed that the effect of the non-orientability was that 
the flux that emanates from one of the ends of the handle becomes reversed 
[Sor]. To a distant observer, Sorkin's non-orientable handle appears to be
a source of net flux. This construction provides a model of individual, 
isolated charges that is based purely upon the topology of space. These ideas 
were further elaborated and extended in [DH], who generalised Sorkin's proposal 
and provided new illustrations of it.

In [Sor, DH], the non-orientability of space was assumed to be concentrated 
in a small region of space, and to be bounded by a sphere $S$. Outside of $S$, 
space remains orientable. These authors showed that there exist field 
configurations where the flux of the magnetic pseudovector field $\bf B$
through $S$ was nonzero, even though $\bf B$ has zero divergence. This was 
interpreted as the presence a new kind of magnetic charge situated inside $\S$. 

Fig.~2 gives a two-dimensional illustration of this idea. A non-orientable 
surface is crossed by magnetic lines of force, and the non-orientability causes 
the lines of force to emerge with the same orientation out of both ends. A 
distant observer would assign the same magnetic charge to both of them. In 
contrast, electric flux lines would have behaved very much as in Fig.~1, and 
each end would have seemed to carry opposite electric charges. This example 
suggests that the non-orientability affects only one kind of charge, namely
the one we called dual or magnetic.

\bigskip
{\centerline{\epsfbox{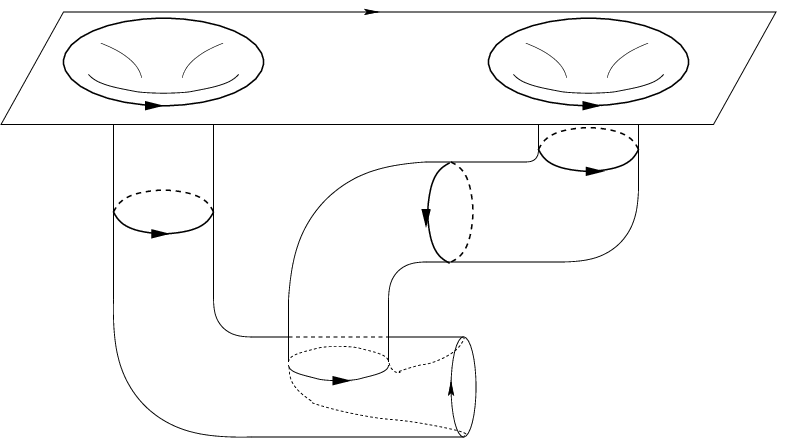}}\nobreak
{\centerline{\fig{A non-orientable handle}}}}
\bigskip

It is not clear if the non-orientable spaces introduced by Sorkin can be
obtained as solutions to General Relativity or, if they could, whether they
would be stable. As with the original idea of Misner and Wheeler, there is no 
scale with which to make sense of the statements that the non-orientable region 
of space is small, and that the observer who measures the flux through $\S$ 
must be far away from that region. In the absence of any known answers to these 
questions, we will ignore them, and focus instead in the purely topological 
aspects of these constructions.

The ideas that will be discussed in the main body of this paper are similar
to those in [Sor], although the point of view will be rather different.
None of the objections pointed out earlier will be answered, and it will not 
be claimed that these models can provide an alternative model of
electromagnetic charge. Rather, the goal will be to show that a different
connection between magnetic charge and non-orientability can be found, in which 
the topological properties of non-orientable spaces appear more explicitly. 

The idea of magnetic charge that we will consider is defined by the flux of 
the electromagnetic field strength $F$ through a sphere $\S$ that bounds a 
non-orientable region of space. Being differential forms, not densities, neither
$F$ nor $dF$ can be integrated on non-orientable spaces. The flux of $F$ through
$\S$ is not problematic, since spheres are always orientable, but the integral 
of $dF$ over the non-orientable interior of $\S$ is not defined. This precludes 
the use of the Stokes theorem for differential forms, and for this reason the
connection between the flux of $F$ and any notion of magnetic charge located
inside $\S$ cannot be of a conventional type. We will propose a way to make 
sense of the Stokes theorem in these circumstances, and show that a notion 
of magnetic charge associated with the flux of $F$ can still be made to work.

Once that is done, we will investigate the topological properties of
non-orientable manifolds that make possible the existence of that kind of 
magnetic charge. It is well known that non-orientable manifolds are 
characterised by the existence of a torsion cycle of codimension one. 
More exactly, if $\N$ is the non-orientable interior of $\S$, then the 
homology group $H_{m-1}(\N,\S)$ contains one and only one torsion subgroup that 
is cyclic of order two. Using that result, we will show that 
the homology class of the torsion cycle determines the value of the magnetic 
charge carried by $\N$, as measured by the flux of $F$ through $\S$.

\section{First notions}
We shall be working in an $(m+1)$-dimensional spacetime of the form 
$\R\times\K$, where the real line $\R$ is the time direction, and the smooth 
$m$-manifold $\K$ represents space. By construction, this spacetime will be 
time-orientable, so it will be orientable if and only if the space manifold 
$\K$ is. In this spacetime there is a closed $(m-1)$-form $F$, the 
electromagnetic field strength. The presence of magnetic or electric sources in 
spacetime is expressed by the dual currents $\mu_m$ and $\mu_e$, which are 
related to the field strength $F$ and its Hodge dual ${*}F$ by the equations,
$$
\eqalign{
dF&=\mu_m,\cr
d{*}F&=\mu_e.\cr}
\eqnu{maxwell}
$$\label\maxwell{\num}\par\noindent
The two $m$-form $\mu_m$ and $\mu_e$ are taken to be purely spatial, \ie\ they 
will have no timelike components. This is a physical assumption based on
the fact that the world-volumes of localised charged sources are always
timelike. The dual current, being transversal to the world-volume, must then
be always spacelike.

Assuming that the Stokes theorem for differential forms can be applied,  
\maxwell\ relate the magnetic and electric charges to the fluxes of $F$ and 
${*}F$ respectively. Suppose that $\N$ is an $m$-dimensional region of $\K$ 
whose frontier is a smoothly embedded $(m-1)$-sphere $\S$. Then the magnetic 
charge contained in $\N$,
$$
Q_m(\N)=\int\limits_\N \mu_m,
\eqnu{mcharge}
$$\label\mcharge{\num}\par\noindent
equals the flux of $F$ through $\S$,
$$
Q_m(\N)=\int\limits_\N \mu_m=\int\limits_\N dF=
\int\limits_\S F.
$$
Similarly, if $\nu$ is a 3-dimensional region of $\K$ bounded by a sphere 
$\sigma$, the electric charge contained in $\nu$, defined as
$$
Q_e(\nu)=\int\limits_\nu \mu_e,
\eqnu{echarge}
$$\label\echarge{\num}\par\noindent
equals the flux of ${*}F$ through $\sigma$,
$$
Q_e(\nu)=\int\limits_\nu \mu_e=\int\limits_\nu d{*}F=
\int\limits_\sigma {*}F.
$$
Although charges can be defined directly in terms of the currents, as in 
\mcharge\ and \echarge, we will only be concerned with those charges that 
can be expressed as fluxes. The general question of whether charges are 
always fluxes or not depends on a rather intrincate interplay between the 
topology of the world-volume of the sources and the topology of the spacetime 
manifold in which they propagate. For a thorough analysis of this question in 
compact orientable manifolds with boundary, see [AO].

\section{Global considerations}

The equations \maxwell, which hold locally at every point in space, leave
open the possibility that $F$ or ${*}F$, or even the dual currents $\mu_m$
and $\mu_e$ may not be globally defined. Suppose that we cover the whole space 
$\K$ with sets $\{U_i\}$ such that both the individual sets $U_i$ and their 
multiple intersections are topologically spheres. An electrically charged 
scalar particle is represented by a complex wave function $\psi(x)$, which can 
be viewed as a vector in an internal two-dimensional space. Given a choice
of orthonormal basis in that internal space, its components $\psi_1(x)$ and 
$\psi_2(x)$ correspond to a definition of real and imaginary parts of 
$\psi(x)$. The internal basis can be chosen in many different ways at every
point $x$, and any two choices are related by an element of the group $O(2)$. 

In general, each set $U_i$ is equipped with its own continuous choice of 
internal bases. If $x$ is a point in the overlap $U_i\cap U_j$, the two 
conventions are related by a transition function $g_{ij}(x)\in O(2)$. This 
situation is indicated by saying that the internal symmetry group $O(2)$ has 
been gauged. 

The reflection element of $O(2)/SO(2)$ reverses the sign of $\psi_2(x)$ and is 
thus the charge conjugation operator. If a transition function $g_{ij}$ 
equals the reflection element, then $\psi(x)$ in $U_i$ becomes its complex 
conjugate $\psi^*(x)$ in $U_j$. Physically, a particle whose electric charge
is positive in the conventions of $U_i$, is regarded as having negative charge
in the conventions of $U_j$. We then say that there is no global notion of 
positive charge. This problem is eliminated in $U_i\cap U_j$ by reversing 
the orientation of the internal frames in, say, $U_j$, in which case 
the transition function becomes an element of the subgroup $SO(2)\subset O(2)$. 
This, however, may only shift the problem over to other intersections. A global
notion of positive charge can always be restored in simply-connected spaces, but
may be impossible in a space that is not simply-connected. When that happens,
the particle could start a travel around some non-contractible cycle of space 
as a positive charge, and return as a negative charge. This is analogous to 
transporting a frame once along a non-contractible loop in a M\"obius band.

In a situation like this, no global definition of electric charge or current is 
possible. Nevertheless, it can be shown that observable quantities, \eg\ the 
electromagnetic forces that acts on an electrically charged particle, are 
independent of these conventions, so that all observers would agree in their 
measurements. Another surprising property of these spaces is that they can hold 
net electric charge even when they are compact, \ie, lacking a boundary 
through which the lines of force could escape. Although it is impossible to 
attribute a definite sign to it globally, the charge will be non-zero
to any observer. These results were first described by Kiskis [Kis], 
who also extended these ideas to non-abelian gauge theories. 

We shall assume that the field strength $(m-1)$-form $F$ is global. Hence, the
above mentioned effects are excluded. Nevertheless, the assumption that
the space manifold is non-orientable has similar consequences for the magnetic
charge as the twist in the internal $O(2)$ space had for the electric charge.
In our model, the dual magnetic current $\mu_m$ is an $m$-form on the 
$m$-manifold $\K$. Consequently, it must be proportional to $\Omega_\K$, the 
volume form of $\K$,
$$
\mu_m=C(w)\Omega_\K,
\eqnu{mucom}
$$\label\mucom{\num}\par
in which $C(w)$ is a distribution localised on $w$, the world-line of the 
magnetic charge. The equation \mucom\ only makes sense globally if $\K$ is 
orientable. If it is not, $\Omega_\K$ can only be defined locally in the sets
$U_i$ that cover the space manifold $\K$. There will then exist a loop 
$\lambda$ along which the $\Omega_\K$ chosen at one point gets transported
back into $-\Omega_\K$, and that would be interpreted as the magnetic charge
having changed sign after traversing $\lambda$. This is a purely classical
effect that resembles the reversal of electric charge that occurs when the 
$O(2)$ bundle that acts on the wave function cannot be reduced to an $SO(2)$ 
bundle [Kis]. 

Because we are assuming that $F$ is global, the same must be true of $\mu_m$. 
In what follows, $\K$ will be non-orientable, and so, as discussed, $\mu_m$ 
can only be defined globally up to a sign. This leaves $\mu_m=0$ as the only 
possible choice, which makes $F$ a closed form, \ie, the first equation in 
\maxwell\ reduces to $dF=0$. It follows from this that no nonzero single-valued 
magnetic charge, as defined in \mcharge, can exist in $\K$. 

Another implication of $\K$ being non-orientable is that, because there is no
global volume element with which to define the Hodge star, the dual field
strength ${*}F$ cannot exist globally. It will reverse sign whenever 
$\Omega_\K$ does, as was the case when it was transported around the loop 
$\lambda$. Therefore $\mu_e$ will also have to change sign upon traversing
$\lambda$. This purely classical effect caused by the non-orientability of 
space resembles the quantum reversal of charge described in [Kis].

In conclusion, whithin the framework of the equations of motion \maxwell, 
spacetimes of the form $\R\times\K$ where the spatial manifold $\K$
is non-orientable cannot support non-zero single-valued electric or magnetic 
currents, and hence neither electric nor magnetic charges in the sense of 
\mcharge\ and \echarge.

It will be shown below that, despite the exclusion of magnetic sources
implied by the condition $dF=0$, it is still possible to define a different 
kind of magnetic charge, understood as a flux of $F$ through a sphere in $\K$, 
which owes its existence precisely to the non-orientability of the space
manifold $\K$.

\section{Fluxes, orientations, and charges}

Given the closed differential $(m-1)$-form $F$ and a smoothly embedded 
$(m-1)$-sphere $S$ in $\K$, the magnetic flux $\Phi(\S)$ is defined as
$$
\Phi(\S)=\int\limits_\S F.
\eqnu{magflux}
$$\label\magflux{\num}\par\noindent
Although $\S$ has been chosen to be a sphere, the definition \magflux\ still
makes sense if $\S$ is allowed to be any orientable $(m-1)$-cycle in $\K$. The
assumption of orientability is needed in order to integrate the differential 
form $F$ on $\S$. In connection to this, it should be pointed out that there 
exists an alternative definition of magnetic flux which involves only densities,
and is thus applicable to any $(m-1)$-cycle in $\K$, orientable or not. The
question arises of whether that alternative definition is equivalent to 
\magflux. It will turn out that the answer depends on whether the interior of 
$\S$ is orientable or not.

In order to explain this other notion of magnetic flux and its relationship 
with \magflux, we temporarily assume that our spacetime $\R\times\K$ is 
orientable, with volume form $\Omega$ and associated Hodge dual $*$. Then the 
magnetic field is defined intrinsically as the vector field $\bf B$ such that
$$
F=dx^0\wedge E+{\rm i}_{\bf B}{\rm i}_0\Omega.
\eqnu{\feb}
$$\label\feb{\num}\par\noindent
The $(m-2)$-form $E$ is the generalisation of the electric field to $(m+1)$ 
spacetime dimensions when $F$ is an $(m-1)$-form, as we are assuming.
The contraction of the timelike unit vector field and the volume form, 
${\rm i}_0\Omega$, coincides whith the spatial volume form $\Omega_\K$ 
introduced in \mucom,
$$
\Omega_\K={\rm i}_0\Omega.
$$
Substituting the decomposition \feb\ into $dF=0$ shows that the divergence of 
$\bf B$ is zero,
$$
\diver{\bf B}=0.
\eqnu{zerodiv}
$$\label\zerodiv{\num}\par\noindent
This result has been obtained under the assumption that the space manifold $\K$
is orientable. Nevertheles, we will keep \zerodiv\ as part of the definition
of a sourceless magnetic field even is space is non-orientable. 

The relationship between $\bf B$ and $F$ is roughly represented by the magnetic
lines of force and their duals shown in Figs.~1 and 2. In those
two-dimensional examples, $\bf B$ is a vector field and $F$ a one-form which 
are represented by the magnetic lines of force and their duals respectively. 
In our framework, $\bf B$ remains a vector field whatever the dimension of 
space, and for that reason the magnetic lines of force are always lines 
in any dimension. In contrast, the rank of $F$ is one less than the
dimension of space, so that the dual lines of force are in general spacelike 
hypersurfaces of codimension one.  

It may be worth pointing out that the intrinsic definition of $\bf B$
given in \feb\ involves a choice of volume form $\Omega$. In particular,
switching $\Omega_\K$ to $-\Omega_\K$ amounts to changing $\bf B$ into
$-\bf B$. This circumstance is familiar from vector calculus, where it appears
as the choice of the ``right-hand rule'' in the definition of the cross product
and the curl. The dependence of $\bf B$ upon the orientation of space is
expressed by saying that $\bf B$ is a {\it pseudovector} field.

Remembering that $\S$ is spacelike, \magflux\ becomes
$$
\Phi(\S)=\int\limits_\S {\rm i}_{\bf B}\Omega_\K.
$$
Because $\S$ is orientable and $\K$ has been assumed, for the time being, to
be orientable as well, it follows that $\S$ is also coorientable. Then the 
volume form $\Omega_\K$ induces a volume form $\Omega_\S$ in $\S$,
$$
\Omega_\S={\rm i}_{\bf n}\Omega_\K
$$
in which $\bf n$ is a unit vector normal to $\S$. Then the flux $\Phi(\S)$
can be rewritten in terms of the pseudovector field $\bf B$ and the unit vector
$\bf n$ as follows,
$$
\eqalign{
\Phi(\S)&=\int\limits_\S {\rm i}_{\bf B}\Omega_\K=
\int\limits_\S {\rm i}_{({\bf B}\cdot{\bf n}){\bf n}} \Omega_\K\cr &=
\int\limits_\S ({\bf B}\cdot{\bf n}){\rm i}_{\bf n} \Omega_\K=
\int\limits_\S ({\bf B}\cdot{\bf n})\Omega_\S.\cr}
$$
The volume element $\Omega_\S$ carries an associated measure in $\S$ that we 
denote by $ds$, and so
$$
\Phi(\S)=\int\limits_\S ({\bf B}\cdot{\bf n})\,ds.
\eqnu{densiflux}
$$\label\densiflux{\num}\par\noindent
Independently of its origin in \magflux, the expression \densiflux\ 
makes sense for any $(m-1)$-cycle $\S$ in $\K$ whether it is orientable or not, 
so long as it possesses the required normal unit vector $\bf n$, that is, so 
long as $\S$ is coorientable in $\K$. This will always be true whenever both 
$\S$ and $\K$ are orientable, as we have been assuming since \feb. Under those 
assumptions, \magflux\ and \densiflux\ are, as we have shown, equivalent. 

We will be mostly interested in the case where $\K$ is non-orientable. Then, 
$\Omega_\K$ can only be defined locally and, through \feb, the same will be 
true of $\bf B$. In this situation, the relationship between \magflux\ and 
\densiflux\ has to be reexamined. It will be useful to keep in mind that 
orientability is an intrinsic property of a space, whereas coorientability is 
not, because it depends on the way it is embedded in a larger space. For 
example, a loop can be embedded in a Klein bottle in two topologically 
inequivalent ways, only one of which makes the loop coorientable, whereas the 
loop itselt is always orientable. In general, an orientable submanifold $\S$ 
embedded in $\K$ is coorientable only if it has an orientable tubular 
neighbourhood [BT]. 

Suppose then that $\S$ is orientable and that it has an orientable tubular 
neighbourhood in $\K$. Then $\Phi(\S)$ can be defined as either the 
flux of $F$ or as the flux of $\bf B$, making use of a volume form defined
in the tubular neighbourhood of $\S$. Nevertheless, because the volume form 
around $\S$ cannot be extended to all of $\K$, the $\bf B$
constructed out of it cannot be extended either. It is not a proper 
divergenceless pseudovector field. In that sense, only \magflux\ provides a 
satisfactory definition of magnetic charge. Alternatively, we could have said 
that $\bf B$ is the fundamental object that exists globally, and that the 
spacelike part of $F$ is constructed out of it once a volume form is chosen. 
From that point of view, $F$ may fail to exist globally in a non-orientable 
space owing to the lack of a globally defined volume form. In that case, the 
correct definition of magnetic charge would have been \densiflux. In short, 
\magflux\ and \densiflux\ cease to be equivalent whenever $F$ and 
$\bf B$ cannot be globally defined simultaneously.

In what follows, we will adopt the view that $F$ is the fundamental object, 
and that it is globally defined unless stated otherwise. Then, \magflux\ will 
be our primary notion of magnetic flux, whereas \densiflux\ will be a secondary 
result, lacking significance if $\bf B$ fails to exist globally.

\subsection{Magnetic charge}

Let us now add the extra assumption that $\S$ separates the space manifold 
$\K$ into two parts that we call $\M$ and $\N$. The part $\M$ will be 
non-compact and orientable, whereas $\N$ is compact but may or may not be
orientable and has $\S$ as its only frontier. An $\N$ non-orientable
will be interpreted below as a quasi-localised region of space whose topology 
is capable of displaying magnetic charge, as first suggested by Sorkin [Sor].
Configurations of this type can be constructed by taking the connected sum of 
two $m$-manifolds, one non-compact and orientable and the other closed and
non-orientable. The connected sum consists in removing the interior of an 
embedded $m$-ball from each manifold, and pasting the remainders together by 
means of an homeomorphism on the boundary spheres of these balls. The two 
boundary spheres become the single sphere $\S$.

By construction, $\S$ will always have an orientable neighbourhood, and hence
will be both orientable and coorientable. Therefore, both definitions 
of magnetic flux, \magflux\ and \densiflux\ are in principle applicable to it. 
In order to emphasise the conceptual difference between the two, we will 
introduce a specific notation for the latter:
$$
\eqalign{
\Phi(\S)&=\int\limits_\S F,\cr
\B(\S)&=\int\limits_\S ({\bf B}\cdot{\bf n})\,ds.\cr}
\eqnu{notats}
$$\label\notats{\num}\par\noindent
If there is net magnetic flux of the first or second kind through $\S$, we will 
say that $\N$ carries magnetic F-charge or B-charge respectively. As has already
been explained, these definitions are only satisfactory when the globally
well-defined object is $F$ or $\bf B$ respectively.

In order to be able to obtain general results concerning the values of 
$\Phi(\S)$ and $B(\S)$, we need to make use of the Stokes theorem, which 
relates those fluxes to the behaviours of $F$ or $\bf B$ in the interior 
of $\N$. In fact, there are two versions of the Stokes theorem, depending on 
whether we are working with differential forms, or with densities [AMR]. 
Integrating differential forms requires an orientation with which to turn the
volume element into a measure. Hence, the Stokes theorem for differential
forms is only applicable to orientable manifolds, and in particular we need
$\N$ to be orientable. If so, then
$$
\int\limits_\N dF=\int\limits_\S F.
\eqnu{stokesa}
$$\label\stokesa{\num}\par\noindent
Remembering that $F$ is exact, it follows that an orientable region $\N$ never
carries magnetic F-charge, 
$$
\N {\hbox{ orientable}}\quad\Longrightarrow\quad \Phi(\S)=0.
\eqnu{noformcharge}
$$\label\noformcharge{\num}\par\noindent
On the other hand, the Stokes theorem for densities is applicable whether $\N$
is orientable or not, because densities can be integrated directly without any
extra requirements. If we depart for a moment from our view that $F$ must 
always be globally defined, and assume that $\bf B$ extends smoothly  
to the interior of $\N$ (at the expense of $F$ if $\N$ is non-orientable), then
$$
\int\limits_\N \diver {\bf B}\, dv=\int\limits_\S ({\bf B}\cdot{\bf n})\,ds,
\eqnu{stokesb}
$$\label\stokesb{\num}\par\noindent
where $dv$ is the measure in the interior of $\N$. But the condition \zerodiv\ 
implies that the region $\N$, orientable or not, never carries magnetic 
B-charge either,
$$
B(\S)=0.
\eqnu{nodensitycharge}
$$\label\nodensitycharge{\num}\par\noindent
In summary, the two notions of magnetic charge defined in \notats\  
agree when $\N$ is orientable, and both vanish. When $\N$ is non-orientable
and $\bf B$ exists globally, then the B-charge is well-defined but also
vanishes, whereas the F-charge is not well-defined. The only remaining
possibility is that $\N$ is still non-orientable, and $F$ globally defined.
The question is then whether the F-charge can ever be non-zero, and what
its values can be.

The Stokes theorem for differential forms \stokesa\ is inapplicable when 
$\N$ is non-orientable, so that no conclusions can be drawn from it. 
Nevertheless, we will find a way to make sense of \stokesa\ for $\N$ 
non-orientable by means of a cell decomposition, as will be explained below. 
This procedure will relate the values of $\Phi(\S)$ to certain topological 
quantities intrinsic to $\N$, independent of the choice of cell decomposition, 
and hence endowed with an invariant meaning.

\subsection{Remarks}

Before explaining how \stokesa\ can be made to work for $\N$ non-orientable,
let us first discuss what could in principle be wrong with it. The right-hand
side still makes sense, because $\S$ is orientable, and so $F$ can be 
integrated on it. In contrast, the left-hand side contains the $m$-form $dF$ 
integrated on the non-orientable $m$-manifold $\N$, an operation that does not
make sense. It is this that makes the Stokes theorem for differential
forms unusable in non-orientable manifolds.

That problem notwithstanding, we must recall that we are not dealing with a
generic differential form, but with a closed one. There is no difficulty 
in integrating $dF=0$ on any $m$-manifold, orientable or not, and so the
left-hand side of \stokesa, being always zero, still makes sense after all. 

Even if each side of \stokesa\ makes sense individually when $dF=0$ whether
$\N$ is orientable or not, the equation itself is only valid for $\N$ 
orientable. Nevertheless, we can think of $\N$ as the union of a number of 
$m$-dimensional cells, and \stokesa\ is applicable in each of the cells. In the 
next section, we will show that this reinterpretation of the manifold as a cell 
complex leads to a version of \stokesa\ that is valid for non-orientable 
manifolds with frontier. Its left-hand side will still be required to be zero, 
but its right-hand side will receive a correction that turns the frontier of 
$\N$ into the more general concept of homology boundary. 

The difference between frontier and homology boundary is fundamental in
what follows. It will be useful to give a short review of these ideas, before
analysing their relevance to the value of the magnetic F-charge.

\section{Two notions of boundary}

In point-set topology, an $m$-manifold with boundary is locally mapped to 
$R_+\times R^{m-1}$, and the boundary is the submanifold that gets mapped 
into $0\times R^{m-1}$. That conforms to the ordinary idea of boundary as a 
frontier where the manifold ends, and in fact we have been referring to this 
notion as ``frontier'' in previous paragraphs. 

From the point of view of homology, an $m$-manifold can be 
regarded as a sum of $m$-cells, each cell being homeomorphic to an $m$-ball. 
The $m$-cells can be visualised as $m$-dimensional polyhedra that fill out 
the entire manifold without overlapping. The boundary of an $m$-cell, 
defined as in point-set topology, \ie, as its frontier, is now regarded as 
the sum of its faces. These faces now are the $(m-1)$-cells in our cell
decomposition of the manifold, and so on. 

The collection of all $k$-cells is known as the $k$-skeleton of the manifold,
where $k$ runs from 0 to $m$. We will adopt the notation $V_a^k$ for the 
elements of the $k$-skeleton, where $a$ is an appropriate index that labels 
its elements. Because the cells $V_a^k$ are homeomorphic to $k$-balls, the 
boundaries $\partial V_a^k$ are homeomorphic to $(k-1)$-spheres.

The precise way in which the boundaries of the $(r+1)$-cells fit into the 
$r$-skeleton is summarised in the incidence matrix $M_{ab}^{r+1,r}$, defined by
$$
\partial V_a^{r+1}=\sum_b M_{ab}^{r+1,r}V_b^r.
\eqnu{incidence}
$$\label\incidence{\num}\par
The entries of these matrices can only be $0$ or $\pm 1$. As a matter of 
terminology, we will say that $V_a^{r+1}$ incides upon $V_b^r$  whenever 
$M_{ab}^{r+1,r}\neq 0$.

The operation of taking the boundary of a cell is required to be nilpotent, 
that is, it must always be the case that $\partial^2=0$. At the level of cells,
this requirement follows from the earlier statement that $\partial V_a^m$ is 
homeomorphic to the sphere $S^{m-1}$, and hence boundaryless. In order to
reflect this nilpotency, the incidence matrices are required to satisfy the 
relationship 
$$
\sum_b M_{ab}^{r+1,r} M_{bc}^{r,r-1}=0.
\eqnu{relmat}
$$\label\relmat{\num}\par
In the definition of the $m$-dimensional cells $V_a^m$, there is implied a 
choice of orientation for each one of them. Recall that the $m$-cells are 
homeomorphic to $m$-balls, and for these the notion of orientation is 
straightforward. Then, just as for $m$-balls, there are two choices possible
for the orientation of each cell. In principle there is no preferred 
way to choose all the orientations. Nevertheless, some choices are
better than others, in a sense to be explained below. For the time being, we 
will assume that every $m$-cell has been given a definite orientation. This
induces an orientation on its boundary in the usual way. Once this is done,
the integration of differential forms is defined both in the interior of the
cells and on their boundaries.

Now, we identify the manifold with the sum of all the $m$-cells, 
$$
\N=\sum_a V_a^m.
\eqnu{oriv}
$$\label\oriv{\num}\par
When we combine this identification between the manifold and its cell
decomposition with the definition \incidence\ of the incidence matrix, the
boundary of $\N$ can be given as 
$$
\partial\N=\sum_a \partial V_a^m =\sum_a\sum_b
M_{ab}^{m,m-1}V_b^{m-1}.
\eqnu{boundsum}
$$\label\boundsum{\num}\par
Every $(m-1)$-cell in the cell decomposition of a manifold is the face of one 
or two $m$-cells. In contrast, cells of dimension $m-2$ and lower can be shared
by any number of higher-dimensional cells, depending on the details of the cell 
decomposition. It follows that the colums of the incidence matrix 
$M_{ab}^{m,m-1}$ can be of three types:
$$
\bordermatrix{
 & V_{a,\ {\rm{type\ I}}}^{m-1}\atop  \cr
 & 0 \cr
 & \cr
 &\vdots \cr
 &  \cr
 & 0 \cr
 & 1 \cr
 & 0 \cr
 & \cr 
 &\vdots  \cr
 & \cr
 & 0  \cr}\qquad{\hbox{or}}\qquad
\bordermatrix{
 & V_{a,\ {\rm{type\ II}}}^{m-1}\atop  \cr
 & 0  \cr
 & \vdots  \cr
 & 0  \cr
 & 1  \cr
 & 0 \cr
 & \vdots  \cr
 & 0  \cr
 & -1 \cr
 & 0  \cr
 & \vdots  \cr
 & 0  \cr}\qquad{\hbox{or}}\qquad
\bordermatrix{
 & V_{a,\ {\rm{type\ III}}}^{m-1}\atop  \cr
 & 0  \cr
 & \vdots  \cr
 & 0  \cr
 & 1  \cr
 & 0  \cr
 & \vdots  \cr
 & 0  \cr
 & 1  \cr
 & 0  \cr
 & \vdots  \cr
 & 0  \cr},
\eqnu{columns}
$$\label\columns{\num}\par
up to overall signs that can be arranged to be as shown. The
first type corresponds to sets $V_{a,\ {\rm{type\ I}}}^{m-1}$ that lie
entirely on the frontier of $\N$. The second and third types
corresponds to cells $V_{a,\ {\rm{type\ II}}}^{m-1}$ or 
$V_{a,\ {\rm{type\ III}}}^{m-1}$ that are shared by two different 
$m$-dimensional cells. The difference between these two types becomes
apparent when we use \boundsum\ to calculate $\partial\N$. All the
type II cells cancel out, and we are left with 
$$
\partial\N=\sum_a V_{a,\ {\rm{type\ I}}}^{m-1}+
2\sum_a V_{a,\ {\rm{type\ III}}}^{m-1}. 
\eqnu{boundtor}
$$\label\boundtor{\num}\par
This formula illustrates the difference between the notions
of boundary used in homology theory and point-set topology. In our notation, 
the latter comprises all the type I cells, and nothing else; as a 
subspace of $\N$, it coincides with the frontier $\S$. The former includes 
also a certain $(m-1)$ cycle formed from all of the type III cells, multiplied
by two. It will be explained below that the appearence of this 
cycle is due to the existence of a torsion subgroup in the homology group
$H_{m-1}(\N,\S;\Z)$ that is cyclic of order two, whenever $\N$ is 
non-orientable.

Lacking a standard terminology, we will use the name ``free boundary''
for the point-set boundary, and the name ``torsion boundary'' for the 
contribution of the torsion $(m-1)$-cycles. The ``homology boundary'' is the 
sum of both. The symbol $\partial\N$ will be reserved for the homology boundary
of $\N$, and so we write
$$
\partial\N=\sum_a V_{a,\ {\rm{type\ I}}}^{m-1}+
2\sum_a V_{a,\ {\rm{type\ III}}}^{m-1}=
(\partial\N)_{\rm free}+(\partial\N)_{\rm torsion}.
\eqnu{defsofbound}
$$\label\defsofbound{\num}\par
As a chain, the free boundary is the sum of all type I cells, and so it 
manifestly depends upon the choice of cell decomposition. As a subspace of 
$\N$, however, it does not, because, as has already been pointed out, it 
coincides with the frontier $\S$. In contrast, the torsion boundary does 
depend on that choice, even as a subspace of $\N$. It will now be shown that, 
if $\N$ is orientable, then there always exists a cell decomposition for which 
there is no torsion boundary. 

\section{Orientability and cell decompositions}

When we calculate $\partial\N$ using an arbitrary cell decomposition, the 
result is generally of the form
$$
\partial\N=(\partial\N)_{\rm free}+2\gamma,
\eqnu{generalform}
$$\label\generalform{\num}\par\noindent
where $\gamma$ is made entirely of the type III cells of dimension $m-1$. 
The torsion boundary is the term $2\gamma$, and $\gamma$ is the torsion cycle. 
This last formula shows that the double cycle $2\gamma$ is homologous to minus 
the free boundary of $\N$. In the language of relative homology, $\gamma$ is a 
relative torsion $(m-1)$-cycle. 

It is known that the torsion subgroup of $H_{m-1}(\N,\S;\Z)$ is zero if 
$\N$ is orientable (see \eg~[Mas]). In that case, $\gamma$ itself must be a 
relative boundary, say $\gamma=\partial\alpha -\beta$ for some chains 
$\alpha\subset\N$ and $\beta\subset\S$. Then,
$$
\partial(\N-2\alpha)=(\partial\N)_{\rm free}-2\beta.
$$
This shows that our cell decomposition had the wrong signs in the
regions $\alpha$ and $\beta$. Reversing those signs provides a corrected cell
decomposition that is more natural in the sense that the $\partial\N$
calculated with its help does not introduce any spureous torsion
boundaries. These are the ``better'' choices we mentioned earlier in
relation to \oriv. 

The situation for non-orientable manifolds is different. No choice of
cell decomposition can eliminate the torsion boundary, although it can
change it by a homology. In these circumstances, it is useful to know that
the frontier of any manifold, orientable or not, is always orientable. Then we 
can begin by orientating the frontier, that is, the type I $(m-1)$-cells, and 
then provide the adjoining $m$-cells with matching orientations. There is no 
natural prescription for orientating the rest of the cells. 

These ideas are best illustrated by means of an example. The M\"obius band 
is a non-orientable 2-manifold with boundary. It can be obtained from the 
2-dimensional projective plane by cutting out an open disk, which leaves behind
a one-dimensional frontier. A cell decomposition of the M\"obius band that only
uses two 2-cells is shown in Fig.~3. An orientation was first chosen for the 
frontier, and then the cells $C_1$ and $C_2$ were orientated acordingly.

\bigskip
{\centerline{\epsfbox{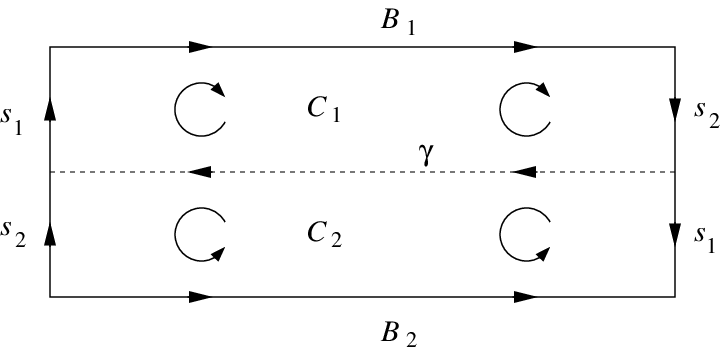}}\nobreak
{\centerline{\fig{A cell decomposition of the M\"obius band}}}}
\bigskip

In the notation shown in the figure, the boundaries of the two cells are
$$
\eqalign{
\partial C_1&=s_1+s_2+B_1+\gamma,\cr
\partial C_2&=-s_1-s_2+B_2+\gamma.\cr}
\eqnu{boundcells}
$$\label\boundcells{\num}\par\noindent
This can be rewritten in terms of an incidence matrix as follows:
$$
\partial\left(\matrix{C_1\cr C_2}\right)=
\left(\matrix{1&1&1&0&1\cr -1&-1&0&1&1}\right)
\left(\matrix{s_1\cr s_2\cr B_1\cr B_2\cr \gamma}\right).
$$
From the incidence matrix we see that $B_1$ and $B_2$ (columns 3 and 4) are 
type I cells, $s_1$ and $s_2$ (columns 1 and 2) are type II, and $\gamma$ 
(column 5) is type III. Then, $B_1$ and $B_2$ form a cell decomposition of the 
frontier of the M\"obius band and $\gamma$ is a torsion cycle. The boundary of 
the M\"obius band is given by  
$$
\partial\N=\partial(C_1+C_2)=B_1+B_2+2\gamma,
\eqnu{boundmoebius}
$$\label\boundmoebius{\num}\par\noindent
which is of the general form \generalform, with $B_1+B_2$ the free boundary 
$\S$, and $2\gamma$ the torsion boundary.

\section{Stokes theorem for cell decompositions}

We are now in a position to apply the Stokes theorem for differential forms 
\stokesa\ to a closed $(m-1)$-form $F$ that exists in a compact $m$-manifold 
$\N$, not necessarily orientable, with non-empty frontier $\S$.
We start by introducing a cell decomposition of the manifold, as in \oriv. 
Because each $m$-cell has been provided with an orientation, \stokesa\ can be 
applied to it,
$$
\int\limits_{V_a^m}dF=\int\limits_{\partial V_a^m}F. 
\eqnu{cellstokes}
$$\label\cellstokes{\num}\par\noindent
Recalling that $F$ is closed, this shows that
$$
\int\limits_{\partial V_a^m}F=0.
\eqnu{cellflux}
$$\label\cellflux{\num}\par\noindent
Let us consider what will happen when the contribution of all the $m$-cells is 
added. Depending on whether the orientations of two adjoining $m$-cells agree
or not, the $(m-1)$-cells that separate them will be type II or III. The 
orientations induced on a type II cell by those of the two $m$-cells that incide
on it will be opposite. Therefore, $F$ is integrated twice on every type II
cell, but with opposite orientations. The two integrals will cancel out, and
so we find that type II cells make no contribution to the sum. In contrast, 
the induced orientations will agree on all type III cells, so that the two 
fluxes will add up instead of cancelling out.

Finally, the contribution of type I cells is straightforward: it is the flux
of $F$ on the free boundary of $\N$, that is, on its frontier $\S$. In fact,
the way the fluxes of $F$ combine when all the contributions from the cells
are added reproduces the calculation of the homology boundary of $\N$, as 
defined in \boundsum,
$$
\sum_a\int\limits_{\partial V_a^m}F=\int\limits_{\partial\N}F.
$$
When this is combined with \cellflux, the result is what we regard as the
Stokes theorem for closed differential forms, in a format that is applicable
whether $\N$ is orientable or not,
$$ 
\int\limits_{\partial\N}F=0.
\eqnu{zeromagflux}
$$\label\zeromagflux{\num}\par\noindent
It has already been explained that, if $\N$ is orientable, then $\partial\N$ 
coincides with its frontier $\S$, in which case \zeromagflux\ is a repetition of
\noformcharge. On the other hand, if $\N$ is non-orientable, $\partial\N$ 
consists of two distinct parts, namely the free and torsion boundaries, as in 
\generalform, and so we find that
$$
\int\limits_\S F=-2\int\limits_\gamma F.
\eqnu{keyresult}
$$\label\keyresult{\num}\par\noindent
The left-hand side of this is $\Phi(\S)$, the magnetic F-charge 
contained inside the sphere $\S$, defined in \notats. The result \keyresult\
shows that the F-charge need not be zero, as we now discuss.

\section{Discussion}

It was found in \nodensitycharge\ that the magnetic B-charge contained in $\N$ 
will always vanish, whether $\N$ is orientable or not, so long as $\bf B$ exists
globally inside $\N$ and remains divergenceless. On the other hand, 
the result \keyresult\ shows that, if $\N$ is non orientable, it may contain a 
non-zero magnetic F-charge.
$$
\Phi(\S)=-2\int\limits_\gamma F.
\eqnu{nonzeroflux}
$$\label\nonzeroflux{\num}\par\noindent
This shows that the presence of the torsion cycle $\gamma$ allows the magnetic 
flux of $F$ through $\S$ to be nonzero. Because the torsion subgrop of 
$H_{m-1}(\N,\S;\Z)$ is cyclic of order two, $\gamma$ will always
exists, and be unique up to homology. In fact, \nonzeroflux\ depends only
on the homology class of $\gamma$, owing to the closedness of $F$,
$$
\int\limits_{\gamma+\partial C} F=\int\limits_\gamma F+
\int\limits_C dF=\int\limits_\gamma F.
$$
This indicates that the cycle $\gamma$, or rather its homology class 
$[\gamma]$, is a peculiar kind of magnetic source. In contrast with a magnetic 
current, which is localised at the world-line of a magnetic charge, the 
homology class $[\gamma]$ is not a localised object, and for that reason it
is not meaningful to ask for a spacetime picture of how exactly the flux 
appears in $\N$ and then crosses the sphere $\S$. A rough representation of
the process can be given only if a choice of $\gamma$ is made, and this is 
illustrated in Fig.~4 for a M\"obius strip. In that figure, the 
frontier of the M\"obius strip was renamed $\S$ in accordance with our 
notation for bounding spheres.

\bigskip
{\centerline{\epsfbox{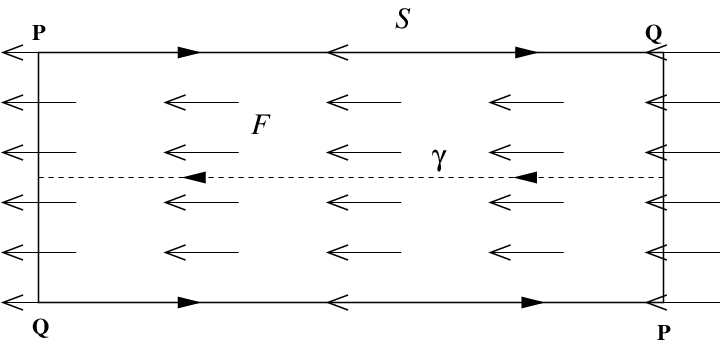}}\nobreak
{\centerline{\fig{A torsion cycle acting as a magnetic source}}}}
\bigskip

A non-zero value for the right-hand side of \nonzeroflux\ would be interpreted 
as $\N$ holding a sort of magnetic charge that does not emanate from 
a localised source, but rather from an intrinsic topological property of
$\N$. Because $F$ is a closed $(m-1)$-form, it represents some cohomology class
in $H^{m-1}(\N;\S)$. Then, a non-zero value for $\Phi(\S)$ requires the 
cohomology group $H^{m-1}(\N;\S)$ to be also non-zero. 

\vfill\eject

\medskip
{\bf{Acknowledgements}}
\smallskip
I am very grateful to David Olive, Paul Martin, Anton Cox, Mark Hadley and 
Allen Hatcher for discussions on different aspects of this work. This research
has been supported by PPARC through the Advanced Fellowship PPA/A/S/1999/00486.
\medskip
{\bf{References}}

[AO] M.~Alvarez, David I. Olive, {\sl Charges and fluxes in Maxwell
theory on compact manifolds with boundary}, submitted to 
Communications in Mathematical Physics; hep-th/0303229

[AMR] R.~Abraham, J.~E.~Marsden, T.~Ratiu, {\sl Manifolds, Tensor Analysis,
and Applications}, Second Edition, Applied Mathematical Sciences 75,
Springer-Verlag New York Inc.~1988

[Bre] Glen E.~Bredon, {\sl Topology and Geometry}, Graduate Texts in 
Mathematics 139, Springer-Verlag New York Inc.~1993

[BG] K. Bronnikov, S. Grinyok, {\sl Charged wormholes with non-minimally
coupled scalar fields. Existence and stability}, contribution to 
Festschrift in honor of Mario Novello; gr-qc/0205131.

[BT] R Bott and L W Tu, {\sl Differential forms in algebraic topology},
Graduate Texts in Mathematics 82, Springer-Verlag New York Inc.~ 1982.

[DH] T.~Diemer, M.~Hadley, {\sl Charge and the topology of spacetime}, 
Classical and Quantum Gravity 16, No 11:3567-3577, 1999; gr-qc/9905069.

[HV] D.~Hochberg, M.~Visser, {\sl Geometric structure of the generic static
traversable wormhole throat}, Physical Review D56:4745-4755, 1997;
gr-qc/9704082.

[Kis] J. Kiskis, {\sl Disconnected gauge groups and the global violation
of charge conservation}, Physical Review D17:3196-3202, 1978

[Lub] E. Lubkin, {\sl Geometric definition of gauge invariance}, 
Annals of Physics 23:233-283, 1963. 

[Mas] William S.~Massey, {\sl A Basic Course in Algebraic Topology}, 
Graduate Texts in Mathematics 127, Springer-Verlag New York Inc.~1991

[Sor] R. Sorkin, {\sl On the relation between charge and topology}, 
Journal of Physics A10:717-725,1977 

[Whe] John A. Wheeler, {\sl Geometrodynamics}, Academic Press, New
York, 1962. 

[WM] Charles W. Misner, John A. Wheeler, 
{\sl Classical physics as geometry: gravitation, electromagnetism, unquantized
charge, and mass as properties of curved empty space}, 
Annals of Physics 2:525-603,1957. 
\end